\documentclass[conference]{IEEEtran}
\IEEEoverridecommandlockouts
\usepackage{cite}
\usepackage{amsmath,amssymb,amsfonts}
\usepackage{algorithmic}
\usepackage{graphicx}
\usepackage{textcomp}
\usepackage{xcolor}

\usepackage{times}
\usepackage{soul}
\usepackage{url}
\usepackage[hidelinks]{hyperref}
\usepackage[utf8]{inputenc}
\usepackage[small]{caption}
\usepackage{amsthm}
\usepackage{booktabs}
\usepackage{color}
\usepackage{enumerate}
\usepackage{algorithm}
\usepackage{bm}
\usepackage{tabularx}
\usepackage{mathrsfs}
\def\BibTeX{{\rm B\kern-.05em{\sc i\kern-.025em b}\kern-.08em
    T\kern-.1667em\lower.7ex\hbox{E}\kern-.125emX}}

\newcommand{\mV}{\mathrm{V}}
\newcommand{\mG}{\mathrm{G}}
\newcommand{\mE}{\mathrm{E}}
\newcommand{\R}{\mathbb{R}}
\newcommand{\Z}{\mathbb{Z}}

\newcommand{\E}{\mathbb{E}}
\newcommand{\betab}{{\bm \beta}}
\newcommand{\xb}{\mathbf{x}}
\newcommand{\yb}{\mathbf{y}}
\newcommand{\Ib}{\mathbf{I}}
\newcommand{\Ab}{\mathbf{A}}
\newcommand{\sbb}{\mathbf{s}}
\newcommand{\1}{\mathbf{1}}
\newcommand{\Mb}{\mathbf{M}}
\newcommand{\Qb}{\mathbf{Q}}
\newcommand{\Lambdab}{{\bm \Lambda}}
\newcommand{\qb}{\mathbf{q}}

\newcommand{\Cb}{\mathbf{C}}
\newcommand{\rb}{\mathbf{r}}
\newcommand{\sigM}{\sigma_{\rm M}}

\newcommand{\MM}{\mathcal{M}}
\newcommand{\db}{\mathbf{d}}

\newcommand{\mR}{\mathrm{R}}
\newcommand{\vb}{\mathbf{v}}
\newcommand{\thetab}{{\bm \theta}}
\newcommand{\pb}{\mathbf{p}}
\newcommand{\Mscr}{\mathscr{M}}

\DeclareMathOperator{\diag}{diag}
\DeclareMathOperator{\re}{Re}
\DeclareMathOperator{\im}{Im}
\DeclareMathOperator{\var}{var}
\DeclareMathOperator{\cov}{cov}
\DeclareMathOperator{\tr}{tr}

\newcommand*\mcup{\mathbin{\mathpalette\mcupinn\relax}}
\newcommand*\mcupinn[2]{\vcenter{\hbox{$\mathsurround=0pt
			\ifx\displaystyle#1\textstyle\else#1\fi\bigcup$}}}

\newtheorem{theorem}{Theorem}
\newtheorem{lemma}{Lemma}

\newtheorem{definition}{Definition}

\newtheorem*{remark}{Remark}

\begin{document}

\title{Distributed Privacy--Preserving Iterative Summation Protocols}

\author{\IEEEauthorblockN{1\textsuperscript{st} Yang Liu}
\IEEEauthorblockA{\textit{Tencent Cloud Product Department} \\
\textit{Tencent}\\
Shenzhen, China \\
clarkieliu@tencent.com}
\and
\IEEEauthorblockN{2\textsuperscript{nd} Qingchen Liu}
\IEEEauthorblockA{\textit{Chair of Information-Oriented Control} \\
	\textit{Technical University of Munich}\\
	Munich, Germany\\
	qingchen.liu@tum.de}
\and
\IEEEauthorblockN{3\textsuperscript{rd} Xiong Zhang}
	\IEEEauthorblockA{\textit{Tencent Cloud Product Department} \\
		\textit{Tencent}\\
		Shenzhen, China \\
		farleyzhang@tencent.com}
\and
\IEEEauthorblockN{4\textsuperscript{th} Shuqi Qin}
\IEEEauthorblockA{\textit{Tencent Cloud Product Department} \\
	\textit{Tencent}\\
	Shenzhen, China \\
	sookieqin@tencent.com}
\and
\IEEEauthorblockN{5\textsuperscript{th} Xiaoping Lei}
\IEEEauthorblockA{\textit{Tencent Cloud Product Department} \\
	\textit{Tencent}\\
	Shenzhen, China \\
	edenlei@tencent.com}
}

\maketitle

\begin{abstract}
In this paper, we study the problem of summation evaluation of secrets. The secrets are distributed over a network of nodes that form a ring graph. Privacy--preserving iterative protocols for computing the sum of the secrets are proposed, which are compatible with node join and leave situations. Theoretic bounds are derived regarding the utility and accuracy, and the proposed protocols are shown to comply with differential privacy requirements. Based on utility, accuracy and privacy, we also provide guidance on  appropriate selections of random noise parameters. Additionally, a few numerical examples that demonstrate their effectiveness are provided.
\end{abstract}


\section{Introduction}

Data mining is a practical technique to extract useful patterns from datasets. Generally, conventional data mining algorithms are developed from a centralized perspective, and effective and robust over an aggregated dataset with an implicit assumption that the collection of such a dataset is  unimpeded. However,
as the information technology rapidly develops, machines that manage to collect and store data, undertake computing and communication tasks are ubiquitous, from powerful servers, to personal agents, such as PCs and smart phones. On the one hand, data collection is usually carried out in a distributed fashion, the data aggregation pushes the communication and computation cost of the central server to a bottleneck on the implementation of centralized data mining methods. On the other hand, the collected data might be sensitive and confidential to the distributed agents, and therefore unanonymized plain data are prohibited to give away directly. Confronted with these challenges in terms of distributed computation and privacy preservation, improvements have to be made on existing data mining methods. The relevant field have drawn much attention \cite{lu2013lightweight}.

The evaluation of sum of a collection of secrets is a fundamental problem that arises in the distributed privacy--preserving data mining. In the field of secure multi--party computation, classical summation protocols are presented \cite{clifton2002tools}, which serve as efficient subroutines for distributed computation. However, the participating parties must follow strict interaction rules in finite steps, leading to inflexibility when facing the disturbance caused by node join and leave. In cryptography, homomorphic encryption and secret sharing can be used as secure summation tools. Nevertheless, the encryption of plaintext yields much more complex results in general, resulting in high communication and computational complexity. Distributed averaging consensus may also be a good approach to evaluating network summation, but the secrets can be perfectly inferred under some observability conditions and the performance against variation of party number is left unexplored \cite{mo2017privacy}. 

In this paper, we propose privacy--preserving summation protocols that are robust against dynamic disturbance and involves no cryptographic encryption. The contributions of this paper are summarized as follows.
\begin{enumerate}[(i)]
	\item Innovative iteration--based distributed privacy--preserving protocols for evaluating summation of secrets are proposed.
	\item Theoretic analysis on utility, accuracy and privacy is provided, illustrated by a few numerical examples.
	\item  The tradeoff problem between utility, accuracy and privacy is solved for providing guidance on the choice of random noise.
\end{enumerate}
The rest of this paper is organized as follows. The problem formulation is introduced in Section \ref{sec:prob_def}. We propose our protocols in Section \ref{sec:the_algorithm}. Theoretic analysis of utility, accuracy and privacy is provided in Section \ref{sec:main_results}. In Section \ref{sec:num_examples}, several examples are give to demonstrate the effectiveness of the proposed protocols. A few concluding remarks are given in Section \ref{sec:conclusions}.

\noindent{\bf Notations.} We let $\R,\R^+$ denote the set of real and positive real numbers, respectively. Similarly, we introduce $\Z,\Z^{\ge0}$ as the set of integers and nonnegative integers, respectively. Let $\E(X)$ denote the expected value of a random variable or vector $X$. We use $f_X(x)$ to denote the probability density function (PDF) of a random variable or vector $X$. Let $\var(\cdot)$ denote the variance of a random variable and $\cov(\cdot)$ denote the covariance of a random vector. We let $\Pr(\cdot)$ denote the probability of an event. We use $\|\cdot\|_{\rm F}$, $\|\cdot\|_p$ to represent the Frobenius norm, $p$--norm of a vector, respectively, and $\|\cdot\|$ to denote the $2$--norm by default. Let $\Ib_n$ denote the identity matrix of size $n$, and $\1_n$ denote the all--ones vector of size $n$. Let $\Mb^\top$ represent the conjugate transpose of a matrix or a vector $\Mb$. Let $\diag(a_1,\dots,a_n)$ denote the diagonal matrix with its diagonal entries being $a_1,\dots,a_n$ from top left to bottom right. In addition, the set of diagonal entries of a real matrix $\Mb$ is denoted as $\diag(\Mb)$. We use $\tr(\cdot)$ to be the trace of a matrix.

\section{Problem Definition}\label{sec:prob_def}

\subsection{Network}

\begin{figure}[h]
	\centering
	\includegraphics[width=5cm]{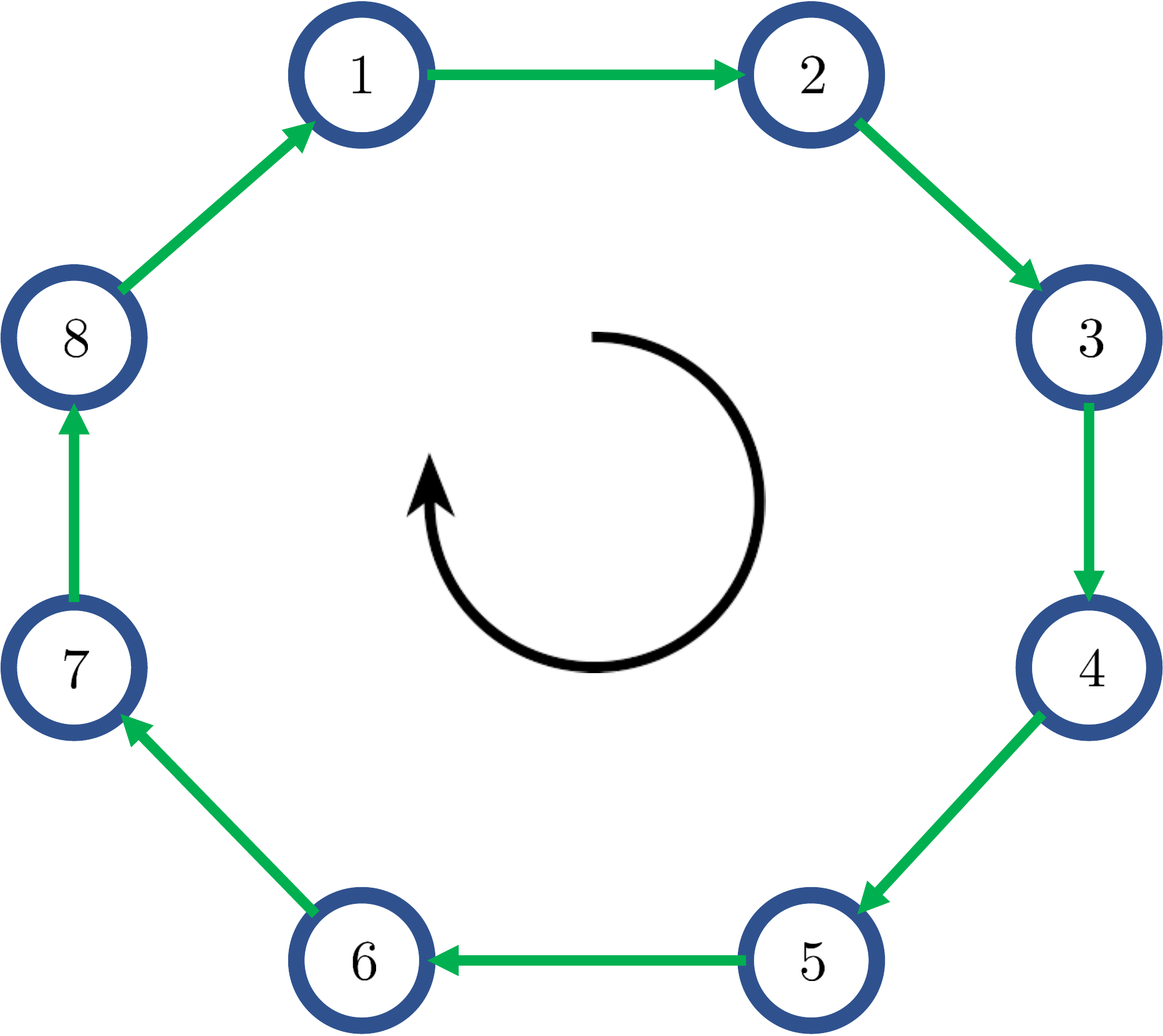}
	\caption{An $8$--node directed ring graph.}
	\label{fig:ring}
\end{figure}

Let a group of nodes be indexed as in $\mV=\{1,\dots,n\}$ with $n>2$. Each node $i\in\mV$ can send information to only one of the other nodes $j\in\mV$, with the connection represented by an ordered pair $(i,j)$.  We suppose, without loss of generality, that node $i$ talks to node $i+1$ for $i=1,\dots,n-1$ and node $n$ talks to node $1$, based on which we introduce a permutation $\pi:\mV\to\mV$ with
\begin{equation}\notag
\pi(i)=\left\{
\begin{aligned}
& i+1&\textnormal{ if }i=1,\dots,n-1;\\
& 1&\textnormal{ otherwise.}
\end{aligned}
\right.
\end{equation}
Then the network can be modelled by a directed ring graph $\mG=(\mV,\mE)$ illustrated in Figure \ref{fig:ring}, where the edge set  $\mE=\{(i,\pi(i)):i\in\mV\}$.

\subsection{General Privacy--Preserving Summation Protocols}\label{sec:principles}

Consider a distributed summation occasion in which each node $i\in\mV$ holds a secret $s_i\in\R$. The collective goal for all nodes is to compute $\sum\limits_{i\in\mV} s_i$ by undertaking certain communication strategies over the network $\mG$, in the meantime each node does not disclose its own secret to others. Formally, we write the principles for Privacy--Preserving Summation Protocols (PPSPs) as follows.

\begin{enumerate}[(i)]
	\item\ [\emph{Communication Principle}]: Each node $i\in\mV$ is only permitted to interact with $\pi(i)$.
	\item\ [\emph{Privacy Principle}] Each node $i\in\mV$ does not disclose $s_i$ to others.
	\item\ [\emph{Utility Principle}] Each node obtains $\sum\limits_{i\in\mV} s_i$ as output.
\end{enumerate}

\section{Achievements with Iterative Method}\label{sec:the_algorithm}

In this section, we introduce realizations of PPSPs. 
We discretize the time as $k=0,1,2,\dots$. Let each node $i\in\mV$ hold a dynamic state $x_i(k)$, initialized as $x_i(0) = s_i$. Introduce $v_i:\Z^{\ge 0}\to\R$ for all $i\in\mV$ satisfying $\lim\limits_{k\to\infty}v_i(k)=0$. Let $\{\beta_i(k)\}_{i\in\mV,k\in\Z^{\ge0}}$ be independent random variables. Let $\pi^{-1}$ denote the inverse of the permutation $\pi$.  Then our deterministic iterative PPSP can be described as follows.

\begin{algorithm}[htb]{{\bf SI-PPSP} Synchronous Iterative PPSP}\\
	\begin{algorithmic}[1]
		\STATE Set $k\gets 0$ and initialize $x_i(0)=s_i$ for all $i\in\mV$.
		\STATE Each node $i$ generates $\beta_i(k)\in\R$ according to a distribution with zero mean and variance $v_i^2(k)$.
		\STATE Each node $i$ sends $d_i(k) = x_i(k) - \beta_i(k)$ to node $\pi(i)$.
		\STATE Each node $i$ updates its state by
		$$x_i(k+1) = \beta_i(k) + d_{\pi^{-1}(i)}(k).$$
		\STATE Set $k\gets k+1$ and go to step 1.
	\end{algorithmic}
\end{algorithm}

We term the described protocol  as a Synchronous Iterative PPSP (SI-PPSP). The ``synchronous" in SI--PPSP is in the sense that  a global network clock  is necessary, in order to schedule all nodes' communication and computation behaviors at each discrete time $k$. It is clear that the communication principle in Section \ref{sec:principles} is met under SI-PPSP. We next show that the privacy and utility principles are also satisfied for the proposed protocol. On the one hand, at each time $k$, the only content that node $i$ sends out is $d_i(k)$ to node $\pi(i)$, which contains not only its current state $x_i(k)$, but a random number $\beta_i(k)$ generated by itself, which prevents node $\pi(i)$ from perfectly learning about its true state. On the other hand, it can be shown by direct computation:
\begin{align*}
\sum\limits_{i\in\mV} x_i(k+1) &=\sum\limits_{i\in\mV} (\beta_i(k) + d_{\pi^{-1}(i)}(k))\\
&= \sum\limits_{i\in\mV} \beta_i(k) + \sum\limits_{i\in\mV} (x_{\pi^{-1}(i)}(k) - \beta_{\pi^{-1}(i)}(k))\\
&= \sum\limits_{i\in\mV} x_i(k).
\end{align*}
This implies that the network sum $\sum\limits_{i\in\mV} x_i(k)$ remains unchanged across the iteration process.

Evidently, a global clock is required within SI-PPSP, which involves a centralized perspective and slightly breaches the distributed setup. Based on \cite{boyd2006randomized}, one can also provide an asynchronized version of SI-PPSP (called AI-PPSP) to facilitate distributed implementations, where each node is equipped with a local Poisson clock and the network demands no central scheduling.



\begin{remark}
	Relevant work is presented in \cite{liu2018gossip1,liu2018gossip2}, where the information exchange rule that preserve network summation is used to shuffle node states, which turns out to be an universal privacy--preserving subroutine for distribution algorithms. In contrast, this paper proposes a sum evaluation protocol by introducing zero--mean random variables with diminishing variance.
\end{remark}

\section{Main Results}\label{sec:main_results}

\subsection{Utility Analysis}

For SI-PPSP, we provide the following result.

\begin{theorem}\label{thm:SI-PPSP_accuracy}
	Consider sequences
	$$\{y_i(k)\}_{k=0}^\infty=\bigg\{\sum\limits_{r=0}^{n-1}x_i(k+r)\bigg\}_{k=0}^\infty$$
	for all $i\in\mV$. Then along SI-PPSP, the following statements hold for each $i\in\mV$.
	\begin{enumerate}[(i)]
		\item If $v_i(k) = \frac{c_i}{k+d_i}$ with $c_i,d_i>0$, then
		$$\lim\limits_{k\to\infty} \E\sum\limits_{i\in\mV}\left|y_i(k) - \sum\limits_{j\in\mV}s_j\right| \le c_M\pi n\sqrt{\frac{n}{6}},$$
		where $c_M=\max\{c_i:i\in\mV\}$.
		\item If $v_i(k) = c_i\phi_i^k$ with $c_i>0$ and $0<\phi_i<1$, then
		$$\lim\limits_{k\to\infty} \E\sum\limits_{i\in\mV}\left|y_i(k) - \sum\limits_{j\in\mV}s_j\right| \le \max\limits_{i\in\mV}c_in\sqrt{\frac{n}{1-\phi_i^2}}.$$
	\end{enumerate}
	
\end{theorem}

The proof of Theorem \ref{thm:SI-PPSP_accuracy} is provided in Appendix A. Evidently, $y_i(k)$ can be termed as a solution estimator. Theorem \ref{thm:SI-PPSP_accuracy} clarifies that each node can add up $n$ consecutive states of its own as an approximation of the network sum, after SI-PPSP executes for a sufficiently long time. Theorem \ref{thm:SI-PPSP_accuracy} provides asymptotic upper bounds for the execution error of SI-PPSP in mean square expectation along time under two commonly used classes of random variance options. This clearly shed theoretic light on the utility of the proposed protocol.

\subsection{Accuracy Analysis}

In the following, we provide a lemma that assists with the proof of further results.
\begin{lemma}\label{lem:1}
	Consider matrices $\Cb^1,\dots,\Cb^m\in\R^{n\times n}$ and random vectors $\rb^1,\dots,\rb^m\in\R^n$. Suppose all the components of $\rb^i,\ i=1,\dots,m$ are pairwise independent. Define $\sigM^i=\max(\diag(\cov(\rb^i)))$. Then
	\begin{equation}\notag
	\tr(\cov(\sum\limits_{i=1}^m \Cb^i\rb^i)) \le \sum\limits_{i=1}^m \|\Cb^i\|_{\rm F}^2\sigM^i.
	\end{equation}
\end{lemma}

The proof of Lemma \ref{lem:1} can be found in Appendix B. The following theorem studies the variance of the solution estimator.
\begin{theorem}\label{thm:converge_in_expectation}
	Consider the same sequences
	$\{y_i(k)\}_{k=0}^\infty, \ i\in\mV$ as in Theorem \ref{thm:SI-PPSP_accuracy}. Then along SI-PPSP, there holds $\lim\limits_{k\to\infty}\E(y_i(k))=\sum\limits_{j\in\mV}s_j$ for all $i\in\mV$. Furthermore, for each $i\in\mV$, the following conclusions can be drawn.
	\begin{enumerate}[(i)]
		\item If $v_i(k) = \frac{c_i}{k+d_i}$ with $c_i,d_i>0$, then
		$$\lim\limits_{k\to\infty}\sum\limits_{i\in\mV}\var(y_i(k))\le\frac{c_M^2\pi^2n^2}{3},$$
		where $c_M=\max\{c_i:i\in\mV\}$.
		\item If $v_i(k) = c_i\phi_i^k$ with $c_i>0$ and $0<\phi_i<1$, then
		$$\lim\limits_{k\to\infty}\sum\limits_{i\in\mV}\var(y_i(k))\le\max\limits_{i\in\mV}\frac{2n^2c_i^2}{1-\phi_i^2}.$$
	\end{enumerate}
\end{theorem}

The proof of Theorem \ref{thm:converge_in_expectation} can be found in Appendix C. Theorem \ref{thm:converge_in_expectation} clarifies the convergence results for the variance of the solution estimator, providing measurement for the accuracy of SI-PPSP.

\subsection{Privacy Analysis}\label{sec:privacy_analysis}
Adversaries against a general privacy--preserving protocol can be simply classified as global and local ones. Global adversaries are usually powerful eavesdroppers, who have access to all communication contents shared among nodes and aim to recover all nodes' secrets based on these observations. In contrast, local adversaries are a subset of protocol participants, who obey the protocol rules but in the meantime try to infer the other nodes' secrets. Privacy analysis against local adversaries are usually termed as semi-honest security in the field of secure multiparty computation. In this section, we will focus on powerful global eavesdroppers for SI-PPSP, the privacy analysis against which will cover semi-honest assumptions.

In practical implementation, SI-PPSP is executed for finite time period $0,1,\dots,K-1$, which is called $K$--step SI-PPSP. Evidently, under SI-PPSP a global adversary aims to infer $\{s_i\}_{i\in\mV}$ based on the observation $\{d_i(k)\}_{i\in\mV,k=0,1,\dots,K-1}$. Such a privacy reconstruction relation can be represented by a mapping $\MM^K:\R^n\to\R^{nK}$, which maps the private data $\{s_i\}_{i\in\mV}$ to the observation $\{d_i(k)\}_{i\in\mV,k=0,1,\dots,K-1}$.
In the following, we formally introduce a few notions that assist with differential privacy analysis \cite{dwork2006calibrating}.

\begin{definition}
	Consider two network secrets $\sbb=[s_1\ \dots\ s_n]^\top$ and $\sbb^\prime=[s_1^\prime\ \dots\ s_n^\prime]^\top$ in vector form. Then $\sbb$ and $\sbb^\prime$ are said to be $\delta$--adjacent if there exists a unique $1\le i \le n$ such that $s_j=s_j^\prime$ for all $j\neq i$ and 
	$$ \left|s_i - s_i^\prime \right| \le \delta. $$
\end{definition}

\begin{definition}
	SI-PPSP preserves $(\epsilon,\delta, K)$--differential privacy if
	$$\Pr(\MM^K(\sbb)\subset\mR)\le e^\epsilon\Pr(\MM^K(\sbb^\prime)\subset\mR)$$
	for all $\mR\in\R^{nK}$ and any two $\delta$--adjacent secrets $\sbb,\sbb^\prime\in\R^n$.
\end{definition}

For SI-PPSP, the following theorem holds.

\begin{theorem}\label{thm:differential_privacy}
	Suppose $\{\beta_i(k)\}_{i\in\mV,k=0,1,\dots,K-1}$ are Laplace distributed random variables.
	\begin{enumerate}[(i)]
		\item If $v_i(k)=\frac{c_i}{k+d_i}$ with $c_i,d_i>0$, then SI-PPSP preserves $$\bigg(c_m^{-1}\delta K\bigg(\frac{K-1}{2} + d_M\bigg),\delta,K\bigg)$$--differential privacy, where $c_m=\min\{c_i:i\in\mV\}$ and $d_M = \max\{d_i:i\in\mV\}$.
		\item If $v_i(k)=c_i\phi_i^k$ with $c_i>0$ and $0<\phi_i<1$, then SI-PPSP preserves $$\bigg(\frac{c_m^{-1}\delta(1-\phi_m^K)}{\phi_m^{K-1}-\phi_m^K},\delta,K\bigg)$$--differential privacy, where $c_m=\min\{c_i:i\in\mV\}$ and $\phi_m=\min\{\phi_i:i\in\mV\}$.
	\end{enumerate}
\end{theorem}

The proof of Theorem \ref{thm:differential_privacy} can be found in Appendix D. Clearly, Theorem \ref{thm:differential_privacy} guarantees that SI-PPSP can preserve differential privacy by only choosing the random noise $\beta_i(k)$ to be Laplace distributed, implying that it complys with state-of-the-art privacy metrics.

\subsection{Utility, Accuracy and Privacy Tradeoff}

The following definition is provided for specifying a class of adversaries.

\begin{definition}
	Adversaries who aim to distinguish $\delta$--adjacent secrets $\sbb,\sbb^\prime\in\R^n$ based on the observation $\MM^K(\sbb)$ and $\MM^K(\sbb^\prime)$ are termed as $\delta$--differential attackers.
\end{definition}

We now provide the following theorem to characterize the tradeoff among utility, accuracy and privacy--preserving capability of SI-PPSP.

\begin{theorem}\label{prop:tradeoff1}
	Consider $K$--step SI-PPSP against $\delta$--differential attackers. Let $\gamma_u,\gamma_a,\gamma_p>0$ be fixed importance balancers for utility, accuracy and privacy. Let $\Mscr_u,\Mscr_a,\Mscr_p$ denote the corresponding metrics concluded in Theorem \ref{thm:SI-PPSP_accuracy}, \ref{thm:converge_in_expectation} and \ref{thm:differential_privacy}.
	Consider the tradeoff problem
	$$\min\limits_{c_i>0,d_i\ge0}\gamma_u\Mscr_u+\gamma_a\Mscr_a+\gamma_p\Mscr_p.$$
	Suppose $v_i(k)=\frac{c_i}{k+d_i}$ with $c_i>0$ and $d_i\ge0$. Then necessarily and sufficiently the optimal $d_1=\dots=d_n=0$, and the optimal choices of $c_i$ are $\bar{c}=c_1=\dots=c_n$ being the real root of the following cubic equation
	\begin{equation}\notag
	4\gamma_a\pi^2n^2\bar{c}^3 + \sqrt{6}\gamma_u\pi n^{\frac{3}{2}}\bar{c}^2- 3\gamma_p\delta K(K-1)=0,
	\end{equation}
	which always uniquely exists.
	
\end{theorem}

The proof of Theorem \ref{prop:tradeoff1} can be found in Appendix E. For the $\frac{1}{k}$--decaying variance case, Theorem \ref{prop:tradeoff1} provides guidance on appropriate selection of the decay parameters by taking into account the utility, accuracy and privacy--preserving capability. However, it is fairly difficult to provide theoretic analysis on the tradeoff problem for the exponentially decaying variance case, because it leads to a complex program and the solution is nontrivial.

\subsection{Node Join and Leave}\label{sec:node_participation_dropout}

Common occasions may occur in which during the execution of PPSP a node chooses to drop out, or an external party would like to join the network sum evaluation. Such node is called a dynamic node. Note that privacy disclosure of the dynamic node's state to all nodes except for the neighbor of the dynamic node is trivial if they know its identity. Therefore, we assume that the dynamic node's identity is anonymous. It turns our that our SI-PPSP and AI-PPSP can perfectly support such disturbance with state update rule altered as follows.

Under SI-PPSP, if a node $i\in\mV$ decides to leave at time $k$, it should send to $\pi(i)$
\begin{equation}\notag
d_i(k) = x_i(k) - s_i,
\end{equation}
inform $\pi^{-1}(i)$ of stopping sending out contents and altering update rule at time $k$ as
\begin{equation}\notag
x_{\pi^{-1}(i)}(k+1) = x_{\pi^{-1}(i)}(k) + d_{\pi^{-1}\circ\pi^{-1}(i)}(k),
\end{equation}
and then leave.
The whole network structure is updated by $\mG^-=(\mV^-,\mE^-)$, where $\mV^-=\mV\setminus\{i\}$ and $\mE^-=\mE\mcup\{(\pi^{-1}(i),\pi(i))\}\setminus\{(\pi^{-1}(i),i),(i,\pi(i))\}$. If an external party $i^+\notin\mV$ is to join, it selects a node $i^\ast\in\mV$ and then directly updates the network by $\mG^+=(\mV^+,\mE^+)$, where $\mV^+=\mV\mcup\{i^+\}$ and $\mE^+=\mE\mcup\{(\pi^{-1}(i^\ast),i^+),(i^+,\pi(i^\ast))\}\setminus\{(\pi^{-1}(i^\ast),\pi(i^\ast))\}$, and the network $\mG^+$ can proceed to employ SI-PPSP.


\section{Numerical Examples}\label{sec:num_examples}

We provide the following example to illustrate the information flow under SI-PPSP.

\medskip

\noindent{\bf Example 1.} We consider a $3$--node ring graph for SI-PPSP. Then along these two protocols, node states evolution process is shown in Table \ref{table:si-ppsp}.
\begin{table}[h!]
	\centering
	\begin{tabularx}{\columnwidth}{X|X|X|X} 
		\hline
		Time $k$ & $x_1(k)$ & $x_2(k)$ & $x_3(k)$ \\ [0.5ex] 
		\hline\hline
		$0$ & $s_1$ & $s_2$ & $s_3$\\
		\hline
		$1$ & $\beta_1(0)+s_3-\beta_3(0)$ & $\beta_2(0)+s_1-\beta_1(0)$ & $\beta_3(0)+s_2-\beta_2(0)$ \\
		\hline
		$2$ & $\beta_1(1)+\beta_3(0)+s_2-\beta_2(0)-\beta_3(1)$ & $\beta_2(1)+\beta_1(0)+s_3-\beta_3(0)-\beta_1(1)$ & $\beta_3(1)+\beta_2(0)+s_1-\beta_1(0)-\beta_2(1)$ \\
		\hline
		$3$ & $\beta_1(2)+\beta_3(1)+\beta_2(0)+s_1-\beta_1(0)-\beta_2(1)-\beta_3(2)$ & $\beta_2(2)+\beta_1(1)+\beta_3(0)+s_2-\beta_2(0)-\beta_3(1)-\beta_1(2)$ & $\beta_3(2)+\beta_2(1)+\beta_1(0)+s_3-\beta_3(0)-\beta_1(1)-\beta_2(2)$ \\
		\hline
		$\vdots$ & $\vdots$ & $\vdots$ & $\vdots$
	\end{tabularx}
	\caption{State evolution under SI-PPSP}
	\label{table:si-ppsp}
\end{table}


\medskip

The following two examples illustrate the implementation performance of SI-PPSP and AI-PPSP, under the involvement of node join and leave.

\medskip

\begin{figure}[h]
	\centering
	\includegraphics[width=9.3cm]{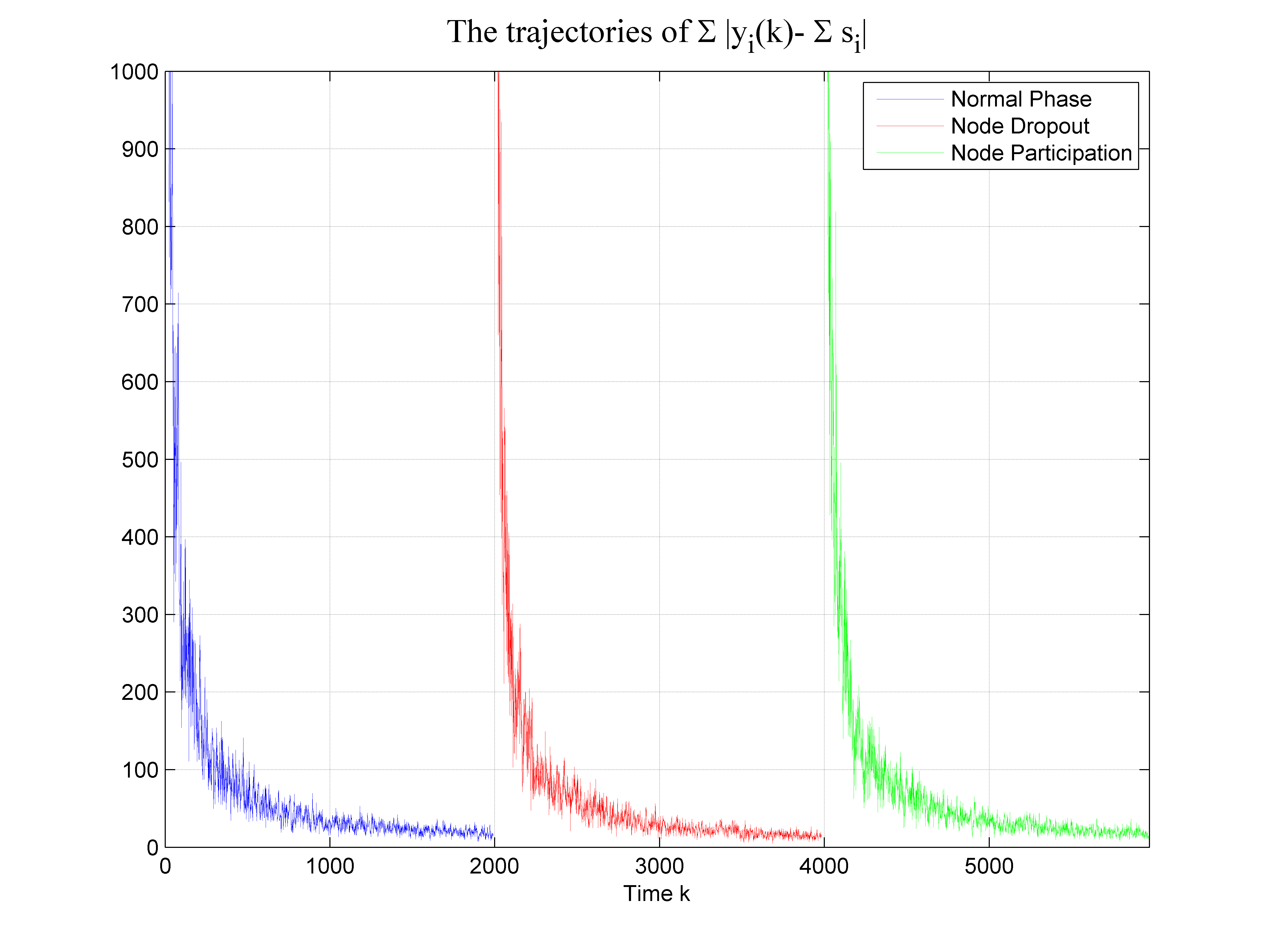}
	\caption{The trajectories of $\sum\left|y_i(k)-\sum s_i\right|$ under SI-PPSP with the involvement of node join and leave, which occur at $k=2000$ and $k=4000$, respectively.}
	\label{fig:si_diff}
\end{figure}

\noindent {\bf Example 2.} Consider a $10$--node ring graph $\mG$. The secrets $s_1,\dots,s_{10}$ held by the nodes are $25.1698, 15.3211, 69.9334, 45.7828, 98.0388,36.6547$, $44.2351, 11.1407, 53.7235, 100$, and it turns out $\sum s_i=500$. Assume $\{\beta_i(k)\}_{i\in\mV,\Z^{\ge 0}}$ are normally distributed. Set $v_i(k)=1000/(k+1)$ for all $i=1,\dots,10$. The experiment of SI-PPSP is conducted in three phases:
\begin{enumerate}[(i)]
	\item At time $k=0,1,\dots,1999$, the network executes SI--PPSP to evaluate the sum $\sum s_i$.
	\item At time $k=2000$, node $n$ chooses to leave and the network follows the method in Section \ref{sec:node_participation_dropout} to overcome the disturbance, and then continue to perform behaviors under SI-PPSP till $k=3999$.
	\item From time $k=4000$, node $n$ decides to join back.
\end{enumerate}
We first plot the trajectories of $y_i(k),\ i=1,\dots,10$ in Figure \ref{fig:ring}. It can be seen that within each phase $p=1,2,3$ each $y_i(k)$ goes to the correct sum
\begin{equation}\notag
\tilde{s}_p=\left\{
\begin{aligned}
& 500&\textnormal{ if }p=1,3;\\
& 400&\textnormal{ if }p=2.
\end{aligned}
\right.
\end{equation}
with tolerable error. We then plot the trajectories of $\sum\limits_{i=1}^{10}\left|y_i(k)-\tilde{s}_p\right|$ for $p=1,2,3$ in Figure \ref{fig:si_diff}. As shown in Figure \ref{fig:si_diff}, affected by the disturbance at $k=2000$ and $k=4000$, node states are forced to deviate from the correct sum $\tilde{s}_p$ for a short period, before going back shortly. The reaction of SI-PPSP against the disturbance clearly shows its robustness under node join and leave.

\begin{figure}[h]
	\centering
	\includegraphics[width=9.3cm]{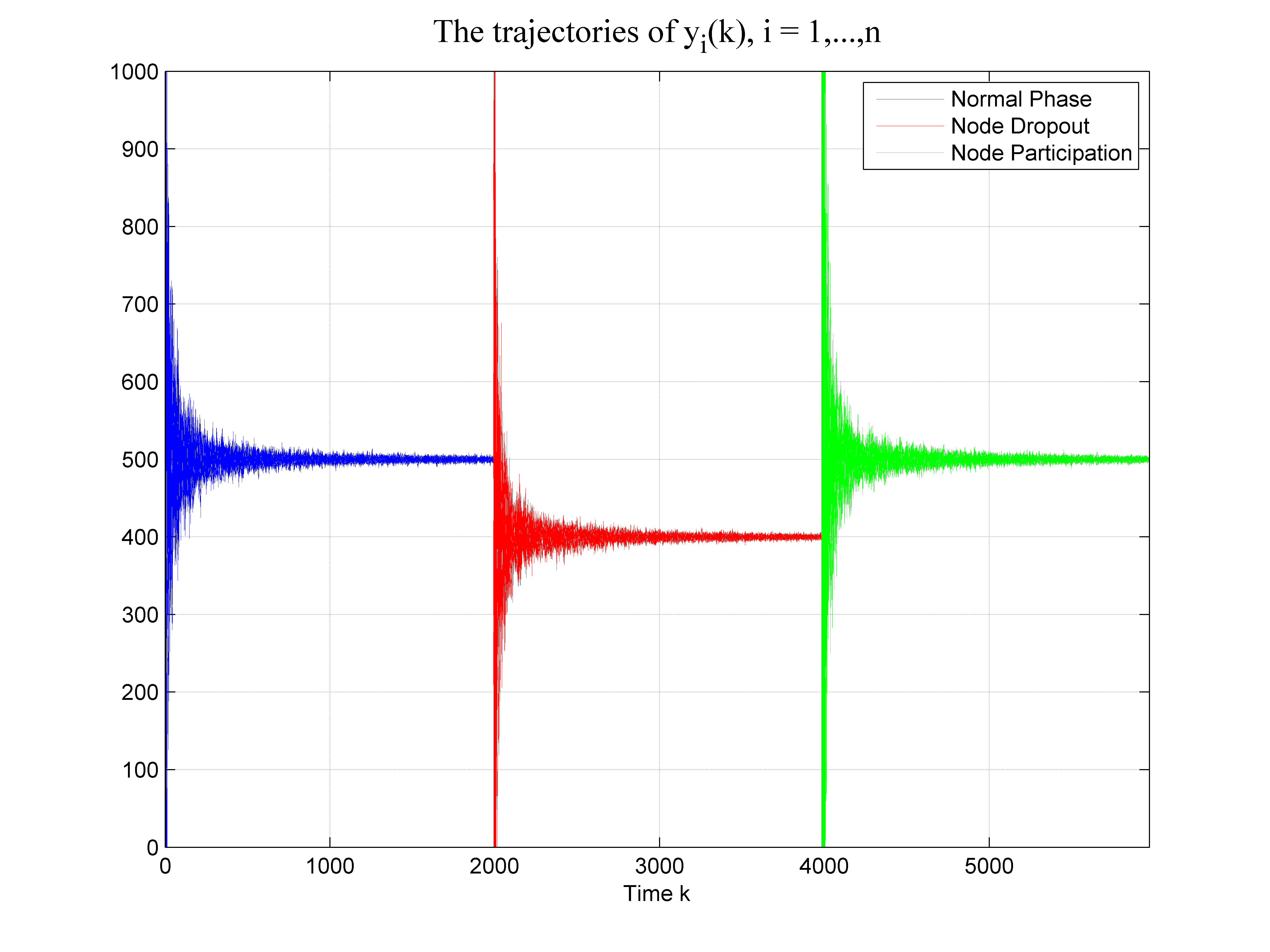}
	\caption{The trajectories of $y_i(k)$s under SI-PPSP with the involvement of node join and leave, which occur at $k=2000$ and $k=4000$, respectively.}
	\label{fig:si_y}
\end{figure}

%
%
%
%
%

\section{Conclusions}\label{sec:conclusions}

This paper studied the problem of evaluating the sum of network secrets. From the dynamic perspective, we proposed distributed privacy--preserving protocols for computing the sum, which are robust against disturbance resulted from node join and leave. Theoretic convergence analysis was provided, demonstrated with several numerical examples that also verifies the effectiveness of the proposed protocols. We also showed that the proposed protocols preserve differential privacy and investigated the tradeoff problem among utility, accuracy and privacy, the solution to which enables us to provide guidance on appropriate selection of random noise parameters.

%

\appendix

\section*{Appendix A. Proof of Theorem \ref{thm:SI-PPSP_accuracy}}

	Introduce
	\begin{align}
	\xb(k) = \begin{bmatrix}
	x_1(k)\\
	\vdots\\
	x_n(k)
	\end{bmatrix},\
	\yb(k) = \begin{bmatrix}
	y_1(k)\\
	\vdots\\
	y_n(k)
	\end{bmatrix},\
	\betab(k) = \begin{bmatrix}
	\beta_1(k)\\
	\vdots\\
	\beta_n(k)
	\end{bmatrix}.\notag
	\end{align}
	We then rewrite node state dynamics of SI-PPSP compactly as
	\begin{align}
	\xb(k+1) = \Ab\xb(k) + (\Ib_n - \Ab)\betab(k),\label{eq:compact_dynamics}
	\end{align}
	where $\Ab\in\R^{n\times n}$ is a circulant matrix in the form
	$$\Ab=\begin{bmatrix}
	0 & 0 & \cdots & 0 & 1\\
	1 & 0 & \cdots & 0 & 0\\
	0 & 1 & \cdots & 0 & 0\\
	\vdots & \vdots & \ddots & 0 & 0\\
	0 & 0 & \cdots & 1 & 0
	\end{bmatrix}.
	$$
	Introduce $\sbb=[s_1\ \dots\ s_n]^\top$ and $s=\1_n^\top\sbb$. Then it is immediate from (\ref{eq:compact_dynamics})
	\begin{equation}\label{eq:dynamics}
	\xb(k) = \Ab^k\sbb + \sum\limits_{m=0}^{k-1}\Ab^m(\Ib_n-\Ab)\betab(k-1-m).
	\end{equation}
	We next study
	\begin{align}
	&\quad\E\|\yb(k)-s\1_n\|^2\notag\\
	&= \E\bigg\|\sum\limits_{r=0}^{n-1}\xb(k+r)-s\1_n\bigg\|^2\notag\\
	&\overset{\rm a)}{=}\E\bigg\|\Ab^k\bigg(\sum\limits_{r=0}^{n-1}\Ab^r\bigg)\sbb-s\1_n\notag\\
	&\quad+\sum\limits_{l=k}^{k+n-1}\sum\limits_{m=0}^{l-1}\Ab^m(\Ib_n-\Ab)\betab(l-1-m)\bigg\|^2\notag\\
	&\overset{\rm b)}{=}\bigg\|\Ab^k\bigg(\sum\limits_{r=0}^{n-1}\Ab^r\bigg)\sbb-s\1_n\bigg\|^2\notag\\ &\quad+\E\bigg\|\sum\limits_{l=k}^{k+n-1}\sum\limits_{m=0}^{l-1}\Ab^m(\Ib_n-\Ab)\betab(l-1-m)\bigg\|^2,\label{eq:two_terms}
	\end{align}
	where a) follows (\ref{eq:dynamics}), and b) is obtained by omitting those terms containing $\E\beta(k)$ due to the zero mean assumption on $\beta_i(k)$s. We now analyze the first term in (\ref{eq:two_terms}). According to \cite{davis2013circulant}, $\Ab$ is diagonalizable as $\Ab=\Qb\Lambdab\Qb^\top$, where the $i$--th column of $\Qb\in\R^{n\times n}$ is given by
	\begin{align*}
	\qb_i&=\frac{1}{\sqrt{n}}
	\begin{bmatrix}
	1 & w_i & w_i^2 & \dots & w_i^{n-1}
	\end{bmatrix}^\top,\\
	w_i&=\exp\bigg(j\frac{2\pi (i-1)}{n}\bigg),
	\end{align*}
	and $\Lambdab = \diag(w_1^{n-1},\dots,w_n^{n-1})$ is a diagonal matrix. Then
	\begin{align}
	&\quad\sum\limits_{r=0}^{n-1}\Ab^r = \Qb\sum\limits_{r=0}^{n-1}\Lambdab^r\Qb^\top\notag\\
	&=
	\Qb
	\begin{bmatrix}
	\sum\limits_{r=0}^{n-1}w_1^{r(n-1)} & & \\
	& \ddots & \\
	& & \sum\limits_{r=0}^{n-1}w_n^{r(n-1)}
	\end{bmatrix}
	\Qb^\top.\label{eq:decom}
	\end{align}
	For $i=2,\dots,n$, direct computation shows if $n$ is even \begin{equation}\notag
	w_i^{r(n-1)}=\left\{
	\begin{aligned}
	& w_i^{(n-r)(n-1)} &\textnormal{ if }  r\neq 0;\\
	& 1&\textnormal{ otherwise,}
	\end{aligned}
	\right.
	\end{equation}
	and if $n$ is odd, then it is satisfied
	\begin{align*}
	\sum\limits_{r=1}^{(n-1)/2}\re(w_i^{r(n-1)})&=\sum\limits_{(n+1)/2}^{n-1}\re(w_i^{r(n-1)})=-\frac{1}{2},\\
	\im(w_i^{r(n-1)})&=\im(w_i^{(n-r)(n-1)}).
	\end{align*}
	Hence, (\ref{eq:decom}) yields
	$\sum\limits_{r=0}^{n-1}\Ab^r = n\qb_1\qb_1^\top$, based on which one has
	\begin{align}
	&\quad\lim\limits_{k=\infty} \bigg\|\Ab^k\bigg(\sum\limits_{r=0}^{n-1}\Ab^r\bigg)\sbb-s\1_n\bigg\|^2\notag\\
	&=\lim\limits_{k=\infty}\bigg\|\begin{bmatrix}
	\qb_1 & \cdots & \qb_n
	\end{bmatrix}
	\begin{bmatrix}
	w_1^{k(n-1)} & & \\
	& \ddots & \\
	& & w_n^{k(n-1)}
	\end{bmatrix}\notag\\
	&\quad\cdot
	\begin{bmatrix}
	n\qb_1^\top\\
	0\\
	\vdots\\
	0
	\end{bmatrix}
	\sbb-s\1_n\bigg\|^2\notag\\
	&=n^2\|\qb_1\qb_1^\top\|^2+s^2\|\1_n\|^2 - 2ns\qb_1^\top\1_n\qb_1^\top\sbb \notag\\
	&= 0.\label{eq:term1}
	\end{align}
	We next investigate the second term in (\ref{eq:two_terms}):
	\begin{align}
	&\quad\lim\limits_{k\to\infty}\E\bigg\|\sum\limits_{l=k}^{k+n-1}\sum\limits_{m=0}^{l-1}\Ab^m(\Ib_n-\Ab)\betab(l-1-m)\bigg\|^2\notag\\
	&\overset{\rm a)}{\le}\lim\limits_{k\to\infty} \sum\limits_{l=k}^{k+n-1}\sum\limits_{m=0}^{l-1}\E\big\|\Ab^m(\Ib_n-\Ab)\betab(l-1-m)\big\|^2\notag\\
	&\overset{\rm b)}{\le}\lim\limits_{k\to\infty}\sum\limits_{l=k}^{k+n-1}\sum\limits_{m=0}^{l-1}\E\|\betab(l-1-m)\|^2\notag\\
	&= \lim\limits_{k\to\infty}\sum\limits_{l=k}^{k+n-1}\sum\limits_{m=0}^{l-1} \sum\limits_{i\in\mV}v_i^2(l-1-m)\notag\\
	&\le n^2\max_{i\in\mV}\sum\limits_{k=0}^\infty v_i^2(k)\notag\\
	&\overset{\rm c)}{\le}\left\{
	\begin{aligned}\label{eq:term2}
	& \frac{c_M^2\pi^2 n^2}{6}&\textnormal{ if }v_i(k) = \frac{c_i}{k+d_i};\\
	& n^2 \max_{i\in\mV}\frac{c_i^2}{1-\phi_i^2}&\textnormal{ if }v_i(k) = c_i\phi_i^k,
	\end{aligned}
	\right.
	\end{align}
	where a) and b) come from norm inequalities
	and c) is from simple calculation of infinite series. Then the proof of the desired statements is completed by (\ref{eq:term1}), (\ref{eq:term1}) and the Cauchy-Schwarz inequality
	\begin{align*}
	\sum\limits_{i\in\mV}\left|y_i(k) - \sum\limits_{j\in\mV}s_j\right|
	&\le\sqrt{\sum\limits_{i\in\mV}\left|y_i(k) - \sum\limits_{j\in\mV}s_j\right|^2}\cdot\sqrt{\sum\limits_{i\in\mV}1^2}\\
	&=\sqrt{n}\|\yb(k)-s\1\|.
	\end{align*}

\section*{Appendix B. Proof of Lemma \ref{lem:1}}

	Let $c_{jl}^i$ denote the $jl$--th entry of $\Cb^i$, and $r_l^i$ denote the $l$--th component of $\rb^i$. Based on the pairwise independence assumption, one can directly write
	\begin{align}
	&\quad\tr(\cov(\sum\limits_{i=1}^m \Cb^i\rb^i))\notag\\
	&= 
	\tr(\cov\begin{bmatrix}
	\sum\limits_{i=1}^m\sum\limits_{l=1}^n c_{1l}^i  r_l^i\\
	\vdots\\
	\sum\limits_{i=1}^m\sum\limits_{l=1}^n c_{nl}^i  r_l^i
	\end{bmatrix})\notag\\
	&= \sum\limits_{j=1}^n \var(\sum\limits_{i=1}^m\sum\limits_{l=1}^n c_{jl}^i  r_l^i)\le \sum\limits_{i=1}^m\sum\limits_{j=1}^n\sum\limits_{l=1}^n (c_{jl}^i)^2  \var(r_l^i)\notag\\
	&\le \sum\limits_{i=1}^m\bigg(\sum\limits_{j=1}^n\sum\limits_{l=1}^n (c_{jl}^i)^2\bigg)\sigM^i = \sum\limits_{i=1}^m \|\Cb^i\|_{\rm F}^2\sigM^i,\notag
	\end{align}
	which completes the proof.

\section*{Appendix C. Proof of Theorem \ref{thm:converge_in_expectation}}

	We will continue to use the notations $\yb(k),\betab(k),\Ab,\Qb,\Lambdab,w_1,\dots,w_n$ in the proof of Theorem \ref{thm:SI-PPSP_accuracy}. Then it follows (\ref{eq:dynamics}) and the analysis in (\ref{eq:term1})
	\begin{align}
	\E\yb(k) &= \Ab^k\bigg(\sum\limits_{r=0}^{n-1}\Ab^r\bigg)\sbb=s\1.\notag
	\end{align}
	Define $\sigM(k)=\max(\diag(\cov(\betab(k))))$. Again based on (\ref{eq:dynamics}), one has using Lemma \ref{lem:1}
	\begin{align}
	&\quad\sum\limits_{i\in\mV}\var(y_i(k))=\tr(\cov(\yb(k)))\notag\\
	&=\tr(\cov(\sum\limits_{l=k}^{k+n-1}\sum\limits_{m=0}^{l-1}\Ab^m(\Ib_n-\Ab)\betab(l-1-m))\notag\\
	&\le \sum\limits_{l=k}^{k+n-1}\sum\limits_{m=0}^{l-1}\|\Ab^m(\Ib_n-\Ab)\|_{\rm F}^2\sigM(l-1-m)\notag\\
	&\le\sum\limits_{l=k}^{k+n-1}\sum\limits_{m=0}^{l-1}\|\Ab\|_{\rm F}^{2m}\|\Ib_n-\Ab\|_{\rm F}^2\sigM(l-1-m) \notag\\
	&= 2n\sum\limits_{l=k}^{k+n-1}\sum\limits_{m=0}^{l-1}\sigM(l-1-m). \label{eq:prop1}
	\end{align}
	Let $k$ go to infinity in (\ref{eq:prop1}). Then one can easily obtain the desired result with simple series computation.


\section*{Appendix D. Proof of Theorem \ref{thm:differential_privacy}}
We will continue to use the following notations in the Appendix A: $\xb(k),\betab(k),\sbb$. Additionally, the superscript symbol prime on a vector $\vb^\prime$ is assumed to apply to each of its components. It is fairly hard to directly study the overall mapping $\MM$. Instead, one can iteratively define a time--varying mapping $\MM_k(\xb(k))=\db(k)$ with $\db(k)=[d_1(k)\ \dots\ d_n(k)]^\top$. Evidently, there holds
$$ \MM^K = \{\MM_k\circ\dots\circ\MM_0(\sbb)\}_{k=1,\dots,K-1}\mcup \MM_0(\sbb). $$
Assuming that each $\MM_k$ preserves $(\epsilon_k,\delta,1)$--differential privacy, according to \cite{mcsherry2009privacy}, $\MM^K$ preserves $\max\bigg\{\sum\limits_{k=0}^l\epsilon_k:l=0,1,\dots,K-1\bigg\}$, namely $\sum\limits_{k=0}^{K-1}\epsilon_k$--differential privacy. The rest of the proof will clarify the way of calculating $\epsilon_k$. For two $\delta$--adjacent secrets $\xb(k),\xb^\prime(k)\in\R$ differing at the $i^\ast$--th component, there holds
\begin{align}
&\quad\frac{\Pr(\MM_k(\xb(k))\subset\mR)}{\Pr(\MM_k(\xb^\prime(k))\subset\mR)}\notag\\
&\overset{\rm a)}{=}\frac{f_{\betab}(\xb(k)-\db(k))}{f_{\betab}(\xb^\prime(k)-\db(k))}\notag\\
&\overset{\rm b)}{=}\exp\bigg(\sum\limits_{i=1}^n \frac{\left|x^\prime_i(k)\right|-\left|x_i(k)\right|}{v_i(k)}\bigg)\notag\\
&\le \exp\bigg(\frac{\left|x_{i^\ast}^\prime(k)-x_{i^\ast}(k)\right|}{v_{i^\ast}(k)}\bigg)\notag\\
&= \exp\big(\delta v_{i^\ast}^{-1}(k)\big),\label{eq:dp1}
\end{align}
where a) is by the definition of probability and b) comes from the fact that $\betab$ is Laplace distributed. We next discuss (\ref{eq:dp1}) in two cases.

\noindent (i) If $v_i(k)=\frac{c_i}{k+d_i}$, then there holds
\begin{equation}\label{eq:dp2}
v_i(k) \ge \frac{c_m}{k+d_M}.
\end{equation}
Then it follows (\ref{eq:dp1}) and (\ref{eq:dp2})
\begin{equation}\label{eq:dp3}
\frac{\Pr(\MM_k(\xb(k))\subset\mR)}{\Pr(\MM_k(\xb^\prime(k))\subset\mR)} \le \exp\big(c_m^{-1}\delta(k+d_M)\big).
\end{equation}
From (\ref{eq:dp3}), one has $\epsilon_k=c_m^{-1}\delta(k+d_M)$, which leads to
\begin{align}
\epsilon = \sum\limits_{k=0}^{K-1} \epsilon_k = c_m^{-1}\delta K(\frac{K-1}{2} + d_M),\notag
\end{align}
which completes the proof of (i).

\noindent (ii) If $v_i(k)=c_i\phi_i^k$, it is analogous $v_i(k)\ge c_m \phi_m^k$. Then
\begin{equation}\notag
\epsilon_k = c_m^{-1}\delta\phi_m^{-k}
\end{equation}
and
\begin{equation}\notag
\epsilon=\sum\limits_{k=0}^{K-1} \epsilon_k = \frac{c_m^{-1}\delta(1-\phi_m^K)}{\phi_m^{K-1}-\phi_m^K}.
\end{equation}
This completes the proof of (ii).

\section*{Appendix E. Proof of Theorem \ref{prop:tradeoff1}}

\noindent (i) Suppose $v_i(k)=\frac{c_i}{k+d_i}$. According to Theorem \ref{thm:SI-PPSP_accuracy}, \ref{thm:converge_in_expectation} and \ref{thm:differential_privacy}, one can summarize the tradeoff among utility, accuracy and privacy as the following optimization problem:
\begin{equation} \label{eq:dp_opt1}
\begin{aligned}
\min_{c_M,c_m>0,d_M\ge0}\qquad & \gamma_u c_M\pi n\sqrt{\frac{n}{6}}+\gamma_a\frac{c_M^2\pi^2n^2}{3}\\
\qquad &+\gamma_p \bigg(c_m^{-1}\delta K\bigg(\frac{K-1}{2} + d_M\bigg)\\
{\rm s.t.} \qquad  &   c_M\ge c_m.
\end{aligned}
\end{equation}
Since the term of $d_M$ in the objective of (\ref{eq:dp_opt1}) is linear and its coefficient $\gamma_p c_m^{-1}\delta K$ is strictly positive, the optimal $d_M$ should be zero independently. Then the optimization problem (\ref{eq:dp_opt1}) can be compactly written as $U(\thetab)$
\begin{equation} \label{eq:dp_opt2}
\begin{aligned}
\min_{\thetab\in\Theta}\qquad & U(\thetab)\\
{\rm s.t.} \qquad  &   \pb^\top
\thetab\le0
\end{aligned}
\end{equation}
where
\begin{align*}
\pb &= \begin{bmatrix}
-1 & 1
\end{bmatrix}^\top,\\
\Theta &= \big\{\begin{bmatrix}
\theta_1&
\theta_2
\end{bmatrix}^\top\in\R^2:\theta_1,\theta_2>0\big\},\\
U(\thetab) &= \gamma_u \theta_1\pi n\sqrt{\frac{n}{6}}+\gamma_a\frac{\theta_1^2\pi^2n^2}{3}\\
&\quad+\gamma_p \theta_2^{-1}\delta K\bigg(\frac{K-1}{2}\bigg).
\end{align*}
Note that the variables $\theta_1$ and $\theta_2$ in (\ref{eq:dp_opt2}) represents $c_M$ and $c_m$, respectively. It can be easily shown that (\ref{eq:dp_opt2}) is a convex optimization problem because $\Theta$ is a convex set, and the objective $U$ and the constraint are both convex functions of $\thetab$. Since the constraint $\pb^\top\thetab$ is an affine function, weak Slater's condition and thus strong duality holds for (\ref{eq:dp_opt2}) according to Section 5.2.3  in \cite{boyd2004convex}.
Therefore, $\thetab^\ast\in\R^2$ is optimal for (\ref{eq:dp_opt2}) if and only if the following Karush–Kuhn–Tucker conditions hold:
\begin{align}
\nabla U(\thetab^\ast) &= - \mu \pb,\label{eq:dp_opt3}\\
\mu \pb^\top\thetab &= 0\label{eq:dp_opt3a}
\end{align}
for some $\mu\ge0$.
Direct computation shows
\begin{align}
\nabla U(\thetab) = \begin{bmatrix}
\frac{2\gamma_a\pi^2n^2}{3}\theta_1 + \gamma_u\pi n\sqrt{\frac{n}{6}}\\
-\frac{\gamma_p\delta K(K-1)}{2\theta_2^2}
\end{bmatrix}.\label{eq:dp_opt4}
\end{align}
Next we study the equation set (\ref{eq:dp_opt3})--(\ref{eq:dp_opt4}). It is evident from (\ref{eq:dp_opt4}) $\nabla U(\thetab^\ast)\neq 0$, and thus by (\ref{eq:dp_opt3}) $\mu\neq 0$. As a result of (\ref{eq:dp_opt3a}), there must hold
\begin{equation}\label{eq:dp_opt5}
\theta_1=\theta_2.
\end{equation}
In addition, by (\ref{eq:dp_opt3}) and (\ref{eq:dp_opt4}), one has
\begin{equation}\label{eq:dp_opt6}
\frac{2\gamma_a\pi^2n^2}{3}\theta_1 + \gamma_u\pi n\sqrt{\frac{n}{6}} = \frac{\gamma_p\delta K(K-1)}{2\theta_2^2}.
\end{equation}
The equation set (\ref{eq:dp_opt5})--(\ref{eq:dp_opt6}) finally leads to a cubic function with respect to $\theta$:
\begin{equation}\label{eq:tradeoff_cubic}
g(\theta)=4\gamma_a\pi^2n^2\theta^3 + \sqrt{6}\gamma_u\pi n^{\frac{3}{2}}\theta^2- 3\gamma_p\delta K(K-1)=0.
\end{equation}
Since $g(0)<0$ and there exists $\hat{\theta}>0$ such that $g(\theta)\ge g(\hat{\theta})\ge 0$ for all $\theta\ge\hat{\theta}$, it can be concluded that there at least exists one $0<\theta^\ast<\hat{\theta}$ such that $g(\theta^\ast)=0$, i.e., $\theta^\ast$ is a positive real root of (\ref{eq:tradeoff_cubic}). Next we show such $\theta^\ast$ is unique. One can directly compute
\begin{align}
g^\prime(\theta)=12\gamma_a\pi^2n^2\theta^2 + 2\sqrt{6}\gamma_u\pi n^{\frac{3}{2}}\theta.\notag
\end{align}
It can be seen that $g^\prime(\theta)>0$ for all $\theta>0$, implying that $g(\theta)$ is strictly increasing over $(0,\infty)$. This results in the uniqueness of $\theta^\ast$ and completes the proof of (i).


\noindent (ii) We now consider the $v_i(k)=c_i\phi_i^k$ case and continue to use the notations $c_M,c_m,\phi_M,\phi_m$ to represent the maximum or minimum of all $c_i$s and $\phi_i$s. To simplify the tradeoff analysis, we slightly loosen the results of Theorem \ref{thm:SI-PPSP_accuracy} and \ref{thm:converge_in_expectation} to
\begin{align}
	\lim\limits_{k\to\infty} \E\sum\limits_{i\in\mV}\left|y_i(k) - \sum\limits_{j\in\mV}s_j\right| &\le c_M n\sqrt{\frac{n}{1-\phi_M^2}},\label{eq:tradeoff_1}\\
	\lim\limits_{k\to\infty}\sum\limits_{i\in\mV}\var(y_i(k))&\le\frac{2n^2c_M^2}{1-\phi_M^2}.\label{eq:tradeoff_2}
\end{align}
Based on (\ref{eq:tradeoff_1}), (\ref{eq:tradeoff_2}) and Theorem \ref{thm:differential_privacy}, one can express the tradeoff problem as
\begin{equation} \label{eq:tradeoff_3}
\begin{aligned}
\min_{c_M,c_m>0,0<\phi_m,\phi_M<1}\qquad & \gamma_uc_M n\sqrt{\frac{n}{1-\phi_M^2}}+\gamma_a\frac{2n^2c_M^2}{1-\phi_M^2}\\
&+\gamma_u\frac{c_m^{-1}\delta(1-\phi_m^K)}{\phi_m^{K-1}-\phi_m^K}\\
{\rm s.t.} \qquad  &   c_M\ge c_m.
\end{aligned}
\end{equation}


\begin{thebibliography}{00}
	

\bibitem{lu2013lightweight}
R. Lu, X. Lin, Z. Shi, and X. Shen,
``A lightweight conditional privacy-preservation protocol for vehicular traffic-monitoring systems,"
IEEE Intelligent Systems, vol. 28, pp. 62--65, 2013.

\bibitem{clifton2002tools}
C. Clifton, M. Kantarcioglu, J. Vaidya, X. Lin, and M. Y. Zhu,
``Tools for privacy preserving distributed data mining,"
ACM Sigkdd Explorations Newsletter, vol. 4, pp. 28--34, 2002.



%



\bibitem{mo2017privacy}
Y. Mo, and R. M. Murray,
``Privacy preserving average consensus,"
IEEE Transactions on Automatic Control, vol. 62, pp. 753--765, 2017.

\bibitem{boyd2006randomized}
S. Boyd, A. Ghosh, B. Prabhakar, and D. Shah,
``Randomized gossip algorithms,"
IEEE/ACM Transactions on Networking, vol. 14, pp. 2508--2530, 2006.

\bibitem{liu2018gossip1}
Y. Liu, J. Wu, R. M. Ian, and G. Shi,
``Gossip algorithms that preserve privacy for distributed computation Part I: the algorithms and convergence conditions,"
IEEE Conference on Decision and Control, pp. 4499--4504, 2018.

\bibitem{liu2018gossip2}
Y. Liu, J. Wu, R. M. Ian, and G. Shi,
``Gossip algorithms that preserve privacy for distributed computation Part II: performance against eavesdroppers,"
IEEE Conference on Decision and Control, pp. 5346--5351, 2018.

\bibitem{dwork2006calibrating}
C. Dwork, F. McSherry, K. Nissim, and A. Smith,
``Calibrating noise to sensitivity in private data analysis,"
Theory of cryptography conference, pp. 265--284, 2006.



\bibitem{davis2013circulant}
P. J. Davis,
``Circulant {M}atrices,"
American Mathematical Soc., 2013.


%
%
%





\bibitem{mcsherry2009privacy}
F. D. Mcsherry,
``Privacy integrated queries: an extensible platform for privacy-preserving data analysis,"
Proceedings of the 2009 ACM SIGMOD International Conference on Management of Data, pp. 19--30, 2009.

\bibitem{boyd2004convex}
S. Boyd, and L. Vandenberghe,
``Convex {O}ptimization,"
Cambridge university press, 2004.

\end{thebibliography}
\end{document}